# Determination of Optimal Size and Number of Movable Energy Resources for Distribution System Resilience Enhancement

Mukesh Gautam*, Eliza Hotchkiss**, and Mohammed Ben-Idris*
*Department of Electrical & Biomedical Engineering
University of Nevada, Reno, Reno, NV 89557, USA
**National Renewable Energy Laboratory, Golden, CO 80401, USA

**SUMMARY**

This paper proposes an approach based on graph theory and combinatorial enumeration for sizing of movable energy resources (MERs) to improve the resilience of the electric power supply. The proposed approach determines the size and number of MERs to be deployed in a distribution system to ensure the quickest possible recovery of the distribution system following an extreme event. The proposed approach starts by generating multiple line outage scenarios based on fragility curves of distribution lines. The generated scenarios are reduced using the k-means method. The distribution network is modeled as a graph where distribution network reconfiguration is performed for each reduced line outage scenario. The combinatorial enumeration technique is used to compute all combinations of total MER by size and number. The expected load curtailment (ELC) corresponding to each locational combination of MERs is determined. The minimum ELCs of all combinations of total MER are used to construct a minimum ELC matrix, which is later utilized to determine optimal size and number of MERs. The proposed approach is validated through a case study performed on a 33-node distribution test system.

mukesh.gautam@nevada.unr.edu

# 1 INTRODUCTION

Extreme events, both natural (e.g., hurricanes, wildfires, ice or hailstorms, earthquakes) and man-made (e.g., cyber- and physical-attacks), have been more frequent over the past decade. For instance, 20 weather-related catastrophes, each costing over $1 billion, occurred in the United States in 2021 alone [1]. Such extreme events have severely damaged crucial power system components, causing prolonged power outages across entire systems. The goal of the electric utilities to provide their consumers with a dependable and resilient electricity supply has been jeopardized by extreme weather events and subsequent outages [2]. To lessen the impact of these events on end customers, post-disaster service restoration (PDSR) procedures must be devised. By making the best use of the resources at hand, PDSR's main objective is to reduce the duration of outages and load curtailments. The most efficient PDSR solutions have been found to be those utilizing smart grid technologies, such as the formation of microgrids, network reconfiguration, repair crew dispatch, distributed generation, energy storage, movable energy resources (MERs), and combinations of these techniques.

MERs are adaptable and transportable resources that can be quickly relocated from staging locations. They may be constructed in a variety of sizes and quickly integrated into the distribution grid following a disaster, thus their adaptability. These resources may be built to handle loads of up to a few megawatts. If no other resources are available, MERs can be deployed to serve local and isolated critical loads when a portion of a distribution system is islanded due to equipment failures or damages.

The deployment of MERs for PDSR has gained significant traction. To increase the resilience of distribution networks, a two-stage robust optimization methodology for scheduling and routing MERs has been developed in [3]. To improve the seismic resilience of distribution systems using MERs, a two-stage PDSR technique, based on mixed-integer linear programming (MILP) has been developed in [4]. For an active distribution system, a MILP-based PDSR technique has been proposed in [5], using an approach to route and schedule energy storage systems for improved resilience. Most of the aforementioned research mainly concentrate on coordinating and dispatching MERs with other PDSR strategies for service restoration, without determining optimal size and number of MERs for a spectrum of potential contingencies.

This paper proposes an approach based on graph theory and combinatorial enumeration for sizing of MERs. High wind speed is taken as an example of weather-related extreme events. A set of multiple line outage scenarios is generated based on forecasted wind speed. Generated scenarios are then reduced using the k-means method. The reduced scenarios are used to determine expected load curtailments (ELCs) when MERs are deployed at various locations for all combinations of total size and number of MERs. The minimum ELCs for each set of MERs are used to construct a minimum ELC matrix, which is utilized to determine optimal size and number of MERs. The proposed approach is validated through a case study on a 33-node distribution test system.

The remainder of the paper is organized as follows: the graph theory modeling of the distribution network is explained in Section 2, the proposed approach and solution algorithm are described in Section 3, a case study on the 33-node system is used to validate the proposed work in Section 4, and Section 5 provides some concluding remarks.

## 2 GRAPH THEORY MODELING OF DISTRIBUTION NETWORK

This section presents the graph theoretic modeling of the distribution network to address the sizing problem being investigated for distribution system resilience. A distribution network forms a meshed network when all the switches (sectionalizing and tie-switches) are closed, and can be represented as an undirected graph $G = (N, E)$, where $N$ is a set of nodes (or vertices) and $E$ is a set of edges (or branches).

### 2.1 Spanning Tree

A subset of the undirected graph (i.e., $G = (N, E)$) known as a spanning tree is one that has the least number of edges connecting each vertex (or node). In a spanning tree, the difference between number of nodes and number of edges is equal to one and all the vertices are connected – there are no cycles (or loops) [6]. Many spanning trees that share the same number of vertices and edges can exist in a linked graph. Each edge of the undirected graph $G$ has a particular value (or weights). Depending on the problem being studied, different edge weights are used. When determining the minimum spanning tree, the total edge weights of the spanning tree are minimized. Figure 1(a) shows a spanning tree of a hypothetical 12-node system. The spanning tree shown in the figure consists of all system nodes (i.e., 12) and 11 closed branches (edges).

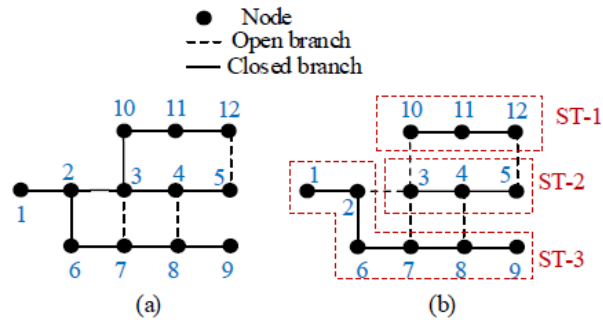

Figure 1: (a) A spanning tree; and (b) a spanning forest for a hypothetical 12-node system

### 2.2 Spanning Forest

In graph theory, a forest is a union of trees that are disconnected from each other. A spanning forest is a forest that spans all vertices of the undirected graph $G$ and is made up of a collection of disconnected spanning trees [7]. Each vertex of the undirected graph $G$ is contained in one of the spanning trees after all of them are connected [8]. When a disconnected graph has many connected components, a spanning forest is formed, which contains a spanning tree of each component [9]. Fig. 1(b) shows the spanning forest formed as a result of disconnection of two additional branches (2–3 and 3–10) in the spanning tree presented in Fig. 1(a). The spanning forest shown in Fig. 1(b) consists of three spanning trees (ST-1, ST-2, and ST-3).

Kruskal's algorithm is used in this work to search for the optimal spanning forest [10]. Based on a given undirected graph, Kruskal's spanning forest search algorithm (KSFSA) first creates a forest $F$ where each graph vertex functions as a single tree. Since KSFSA is a greedy algorithm, it continues, at each iteration, to connect the subsequent least-weight edge that avoids a loop or cycle to the forest $F$. The optimal spanning forest is the forest $F$ that is created after the final iteration. Fig. 2 shows the flowchart of KSFSA.

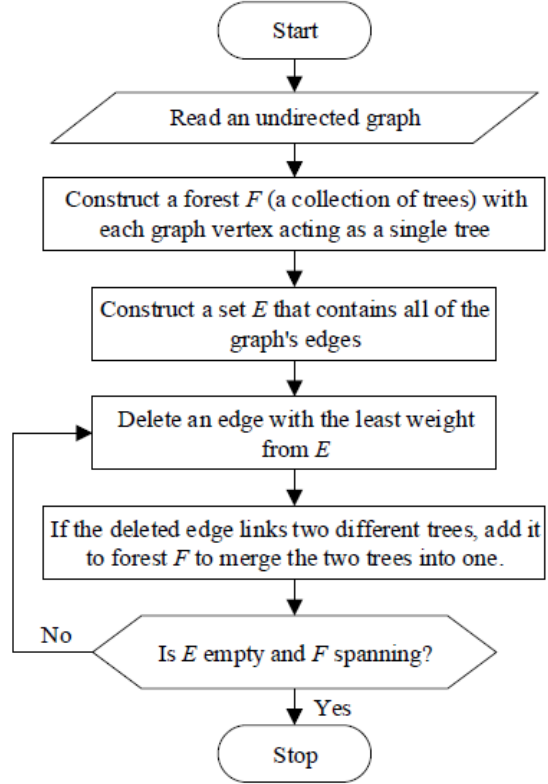

Figure 2: Flowchart of Kruskal's spanning forest search algorithm

## 3 PROPOSED APPROACH

This section presents the proposed approach for event modeling, scenario generation and reduction, and construction of a minimum ELC matrix.

### 3.1 Extreme Event Modeling and Scenario Generation

In this work, the weather-related fragility curve is used to model extreme events and generate multiple line outage scenarios. A fragility curve is applied to characterize the performance and vulnerabilities of different system components confronting uncertain weather-related extreme events. The failure probabilities of each component are obtained by mapping the weather forecast and monitoring data to the fragility curve [11]. We have taken the multiple line outages caused by high wind speeds as an example of a weather-related extreme event in this study. Mathematically, the probability of line outages caused by high wind speeds can be represented as follows:

$$P_l(\omega) = \begin{cases} \overline{P_l}, if\ \omega < \omega_{crl} \\ P_{l.hw}(\omega), if\ \omega_{crl} \leq \omega < \omega_{cpse} \\ 1, if\ \omega \geq \omega_{cpse} \end{cases} \quad (1)$$

where $P_l$ is the probability of line failure as a function of wind speed, $\omega$; $\overline{P_l}$ is the failure probability at normal weather condition; $P_{l.hw}$ is the probability of line failure at high wind; $\omega_{crl}$ is the critical wind speed (i.e., the speed above which the distribution lines start experiencing failure); and $\omega_{cpse}$ is the speed above which the distribution lines completely collapse [12].

## 3.2 Scenario Reduction Using k-means Method

The accuracy of an approach is always improved when a large number of line outage scenarios is used. However, solving the problem with many scenarios takes more time. The generated line outage scenarios are, therefore, reduced using the k-means method in this work to make the proposed approach computationally tractable. The k-means method is an iterative procedure that attempts to split a set of scenarios into a set of unique clusters. It attempts to minimize distance between scenarios in the same cluster while maximizing the distance between different clusters. In addition, when scenarios are assigned to a cluster, the distance between them and the cluster centers is kept to a minimum [13].

## 3.3 Computing Expected Load Curtailment (ELC)

The ELC for $i^{th}$ combination of MER locations is determined using the amount of curtailed critical load for each reduced line outage scenario as follows.

$$ELC_i = \sum_{j=1}^{K} Pr(j) \times LC_i(j), \qquad (2)$$

where $K$ is the total number of reduced scenarios; $Pr(j)$ is the probability of the $j^{th}$ reduced scenario; and $LC_i(j)$ is the critical load curtailment of the $j^{th}$ reduced scenario for the $i^{th}$ combination of MER deployment locations, which is calculated as follows.

$$LC_i(j) = \sum_{m=1}^{N} \omega_m \Delta P_{mi}(j), \qquad (3)$$

where $\Delta P_{mi}(j)$ is the load curtailment at node $m$ of the $j^{th}$ reduced scenario for the $i^{th}$ combination of MER deployment location(s); $\omega_m$ is the critical load factor at node $\omega$; and $N$ is the total number of nodes in the system. While computing the critical load curtailment, the nodal power balance constraints and radiality constraint should always be satisfied.

The proposed methodology to determine optimal size and location of MERs can be summarized in the following steps.
1) Collect system data, including generation data, line data, load data, etc.
2) Generate a set of multiple line outage scenarios based on weather forecasting and monitoring data.
3) Generate a set of reduced scenarios along with their probabilities using a scenario reduction technique.
4) Enumerate all combinations of total size and number of MERs, based on a certain level of granularity.
5) For each combination of total size and number of MERs, determine ELCs corresponding to each combination of MER deployment location.
6) Determine minimum ELC for each combination of total size and number of MERs.
7) Construct a matrix of minimum ELCs based on a combination of total size and number of MERs.
8) Calculate the derivative of the minimum ELC matrix with respect to the number of MERs, then convert the derivative matrix into a vector by averaging over total sizes of MERs. Note that the optimal number of MERs is the number corresponding to the entry of the vector whose sign changes from negative to positive.

9) Calculate the second derivative of the minimum ELC matrix with respect to size of MERs. Convert the second derivative matrix into a vector by averaging over numbers of MERs. Here, the optimal size of MERs is the size corresponding to the maximum entry of the vector.

The flowchart of the proposed approach, following these nine steps to construct the minimum ELC matrix, is shown in Figure 3.

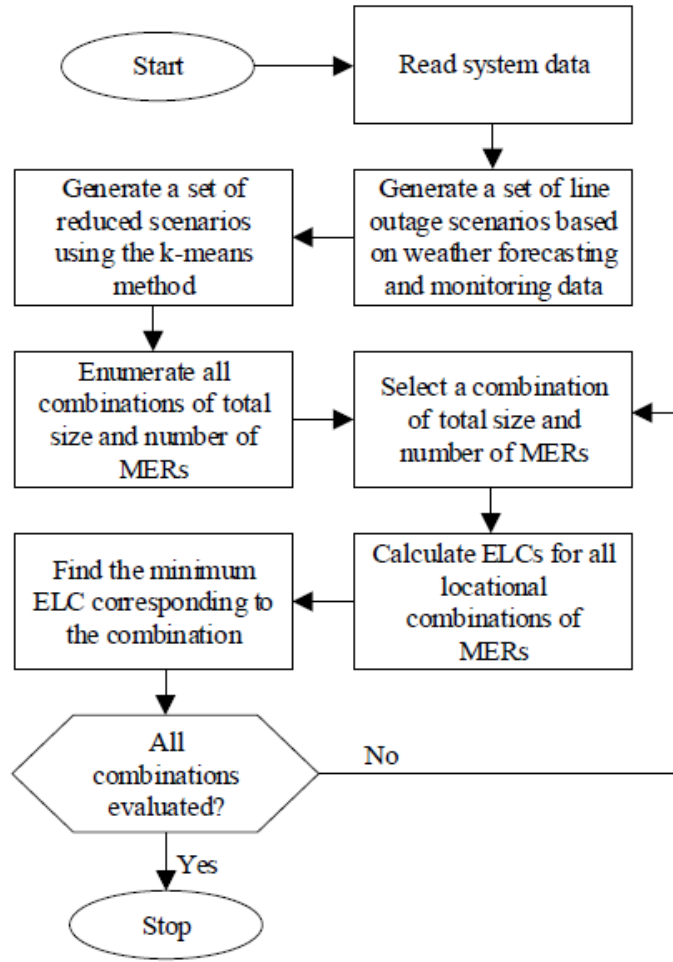

Figure 3: Flowchart of the proposed approach to construct the minimum ELC matrix

## 4 CASE STUDY AND DISCUSSION

A case study was created to test the proposed approach and is described in this section, along with an implementation and results discussion.

### 4.1 System Description

To demonstrate the effectiveness of the proposed approach, a 33-node system is used for numerical simulations. A 33-node distribution test system is a radial distribution system with 33 nodes, 32 branches, and 5 tie-lines (37 branches) [14]. All branches, including tie-lines, are numbered from 1 to 37. The system's overall load is 3.71 MW, and the critical loads are assigned at each node, in kW. The locations and amounts of critical loads considered for the 33-node system are shown in Table 1.

Table 1: Location of Critical Loads for the 33-node system

| Nodes | Critical Loads (kW) | Nodes | Critical Loads (kW) |
|---|---|---|---|
| 4 | 60 | 20 | 45 |
| 5 | 30 | 21 | 45 |
| 6 | 60 | 22 | 45 |
| 7 | 200 | 23 | 45 |
| 8 | 200 | 26 | 60 |
| 9 | 60 | 27 | 60 |
| 10 | 30 | 28 | 60 |
| 11 | 25 | 29 | 60 |
| 18 | 45 | 30 | 60 |
| 19 | 45 | 33 | 30 |

4.2 **Implementation and Results**

For the implementation of the proposed approach, multiple line outage scenarios are generated by considering a high wind speed event as an example of a weather-related extreme event. The critical wind speed of 30 m/s, and the collapse speed of 55 m/s, are assumed for the fragility model (1) under consideration [12]. The failure probability of 0.01 is considered at normal weather conditions. The failure probability starts increasing after 30 m/s and varies linearly up to 55 m/s. The wind fragility curve for distribution lines is shown in Fig. 4. In this modeling, 10,000 random outage scenarios are generated, and the k-means method is used to reduce the generated scenarios into 200 reduced outage scenarios for wind speeds of 38 m/s. The k-means method results in 200 reduced line outage scenarios, along with their failure probabilities.

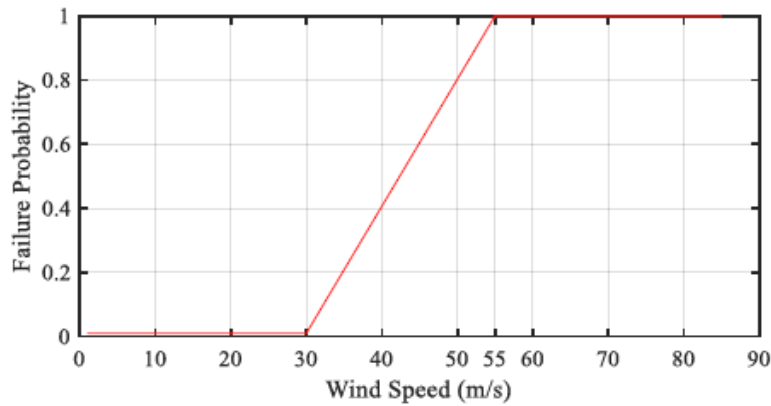

Figure 4: Wind fragility curve for distribution lines

The combinatorial enumeration technique is used to compute all combinations of total size and number of MERs. The total sizes of MERs ranging from 500 kW to 1900 kW are taken at a granularity level of 100 kW. For each total MER size, the number of MERs ranging from 1 to 10 are taken. For combinations consisting of a single MER, its size is equal to the total MER size. For combinations with multiple MERs, the individual sizes of MERs are assumed to be equal to the total MER size divided by the number of MERs. For each combination of total MER size and the number of MERs, there are multiple locational combinations. For each locational combination of MERs, the expected load curtailment (ELC) is determined by considering critical load curtailments corresponding to each of the 200 reduced line outage scenarios and their failure probabilities.

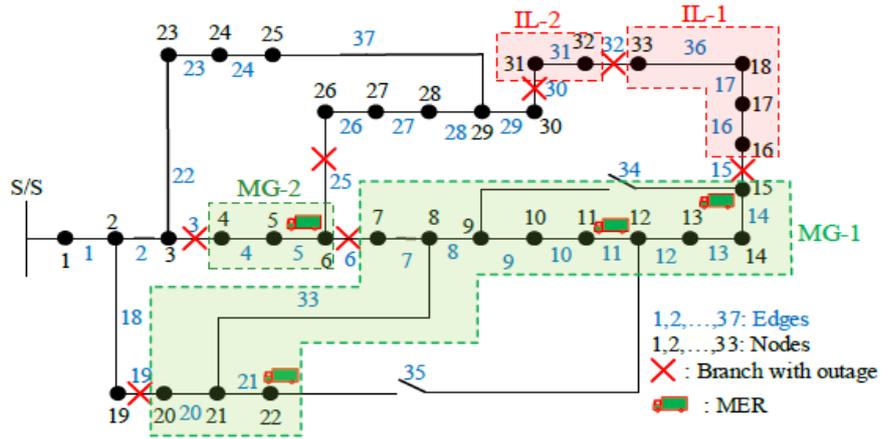

Figure 5: A test case for the 33-node distribution system

Fig. 5 shows the case for a reduced scenario where outages of lines 3, 6, 15, 19, 25, 30, and 32 occur. In this scenario, the distribution network is reconfigured by closing tie-switches 33, 36, and 37, using KSFSA. The tie-switches 34 and 35 are open to maintain radial configuration. When MERs with 300 kW capacity each are deployed at nodes 6, 11, 15, and 22, two microgrids (MG-1 and MG-2) and two isolates (IL-1 and IL-2) are formed. The two isolates are devoid of power supply. The total critical loads of IL-1 and IL-2 are 0 kW and 75 kW, respectively. Therefore, the total critical load curtailment for this reduced scenario is 75 kW. When the procedure shown in Figure 3 is followed, the minimum ELC matrix is formed, as shown in Table 2.

Table 2: Minimum ELC matrix

| Total MER size (kW) /Number of MERs | 1 | 2 | 3 | 4 | 5 | 6 | 7 | 8 | 9 | 10 |
|---|---|---|---|---|---|---|---|---|---|---|
| 500 | 306.633 | 297.857 | 293.235 | 291.224 | 290.514 | 289.6 | 289.255 | 289.167 | 289.376 | 290.089 |
| 600 | 275.965 | 263.233 | 257.222 | 255.605 | 254.325 | 253.815 | 252.984 | 252.634 | 252.506 | 252.253 |
| 700 | 247.683 | 229.533 | 223.513 | 219.461 | 218.463 | 218.11 | 217.191 | 216.653 | 216.266 | 216.003 |
| 800 | 221.297 | 196.57 | 189.77 | 183.764 | 183.924 | 182.204 | 181.789 | 181.035 | 180.408 | 180.448 |
| 900 | 196.881 | 166.573 | 157.104 | 151.897 | 150.232 | 147.759 | 146.951 | 146.8 | 145.387 | 146.488 |
| 1000 | 176.251 | 141.16 | 128.767 | 124.478 | 120.322 | 117.259 | 116.224 | 115.211 | 114.68 | 117.174 |
| 1100 | 158.803 | 118.702 | 103.926 | 99.6455 | 94.373 | 89.4748 | 89.5464 | 87.0585 | 88.4829 | 91.4035 |
| 1200 | 147.327 | 102.806 | 85.6925 | 78.973 | 74.4985 | 67.6935 | 68.5553 | 65.301 | 67.0478 | 70.961 |
| 1300 | 142.561 | 94.277 | 75.9558 | 65.8495 | 62.1615 | 54.7228 | 54.7878 | 53.2583 | 56.2931 | 59.176 |
| 1400 | 142.561 | 91.294 | 72.5542 | 59.021 | 56.0005 | 50.5713 | 49.412 | 49.463 | 52.2787 | 54.57 |
| 1500 | 142.561 | 88.877 | 70.33 | 54.7625 | 52.3145 | 47.8405 | 46.2631 | 46.8567 | 48.994 | 51.5195 |
| 1600 | 142.561 | 86.918 | 69.044 | 52.022 | 49.7855 | 46.564 | 44.1306 | 45.147 | 46.931 | 49.526 |
| 1700 | 142.561 | 85.248 | 67.8728 | 50.234 | 48.056 | 45.3958 | 42.9269 | 44.2593 | 45.3977 | 47.774 |
| 1800 | 142.561 | 83.863 | 65.9125 | 49.1715 | 46.884 | 44.6805 | 42.3654 | 43.2615 | 44.36 | 46.214 |
| 1900 | 142.561 | 82.862 | 64.5482 | 48.609 | 45.8485 | 44.1748 | 41.9911 | 42.5153 | 43.6782 | 44.897 |

In general, when the total MER size or the number of MERs increases, the minimum ELC continues to decrease, as shown in Table 2. However, the rate of change of minimum ELCs is not the same when the total MER size or the number of MERs increases. Due to this reason, the derivatives of the matrix are computed to analyze the rate of change of minimum ELCs.

Table 3: The first derivative of the Minimum ELC matrix w.r.t. number of MERs

| Total MER size (kW) /Number of MERs | 1 | 2 | 3 | 4 | 5 | 6 | 7 | 8 | 9 | 10 |
|---|---|---|---|---|---|---|---|---|---|---|
| 500 | -8.78 | -6.70 | -3.32 | -1.36 | -0.81 | -0.63 | -0.22 | 0.06 | 0.46 | 0.71 |
| 600 | -12.73 | -9.37 | -3.81 | -1.45 | -0.90 | -0.67 | -0.59 | -0.24 | -0.19 | -0.25 |
| 700 | -18.15 | -12.08 | -5.04 | -2.53 | -0.68 | -0.64 | -0.73 | -0.46 | -0.33 | -0.26 |
| 800 | -24.73 | -15.76 | -6.40 | -2.92 | -0.78 | -1.07 | -0.58 | -0.69 | -0.29 | 0.04 |
| 900 | -30.31 | -19.89 | -7.34 | -3.44 | -2.07 | -1.64 | -0.48 | -0.78 | -0.16 | 1.10 |
| 1000 | -35.09 | -23.74 | -8.34 | -4.22 | -3.61 | -2.05 | -1.02 | -0.77 | 0.98 | 2.49 |
| 1100 | -40.10 | -27.44 | -9.53 | -4.78 | -5.09 | -2.41 | -1.21 | -0.53 | 2.17 | 2.92 |
| 1200 | -44.52 | -30.82 | -11.92 | -5.60 | -5.64 | -2.97 | -1.20 | -0.75 | 2.83 | 3.91 |
| 1300 | -48.28 | -33.30 | -14.21 | -6.90 | -5.56 | -3.69 | -0.73 | 0.75 | 2.96 | 2.88 |
| 1400 | -51.27 | -35.00 | -16.14 | -8.28 | -4.22 | -3.29 | -0.55 | 1.43 | 2.55 | 2.29 |
| 1500 | -53.68 | -36.12 | -17.06 | -9.01 | -3.46 | -3.03 | -0.49 | 1.37 | 2.33 | 2.53 |
| 1600 | -55.64 | -36.76 | -17.45 | -9.63 | -2.73 | -2.83 | -0.71 | 1.40 | 2.19 | 2.60 |
| 1700 | -57.31 | -37.34 | -17.51 | -9.91 | -2.42 | -2.56 | -0.57 | 1.24 | 1.76 | 2.38 |
| 1800 | -58.70 | -38.32 | -17.35 | -9.51 | -2.25 | -2.26 | -0.71 | 1.00 | 1.48 | 1.85 |
| 1900 | -59.70 | -39.01 | -17.13 | -9.35 | -2.22 | -1.93 | -0.83 | 0.84 | 1.19 | 1.22 |
| Average over total MER sizes | -39.93 | -26.78 | -11.50 | -5.92 | -2.83 | -2.11 | -0.71 | 0.26 | 1.33 | 1.76 |

Table 4: The second derivative of the Minimum ELC matrix w.r.t. total size of MERs

| Total MER size (kW) /Number of MERs | 1 | 2 | 3 | 4 | 5 | 6 | 7 | 8 | 9 | 10 | Average over number of MERs |
|---|---|---|---|---|---|---|---|---|---|---|---|
| 500 | 1.19 | 0.46 | 1.15 | -0.26 | 0.16 | 0.04 | 0.24 | 0.28 | 0.31 | 0.79 | 0.44 |
| 600 | 1.67 | 0.65 | 1.14 | -0.15 | 0.49 | -0.01 | 0.34 | 0.37 | 0.41 | 0.97 | 0.59 |
| 700 | 2.04 | 1.34 | 0.83 | 1.05 | 0.96 | 0.28 | 0.46 | 0.67 | 0.56 | 1.14 | 0.93 |
| 800 | 2.41 | 2.81 | 1.61 | 3.14 | 1.70 | 1.67 | 1.41 | 1.44 | 1.59 | 2.13 | 1.99 |
| 900 | 3.18 | 3.77 | 3.31 | 3.83 | 3.09 | 3.02 | 3.21 | 2.53 | 3.49 | 3.61 | 3.30 |
| 1000 | 4.03 | 4.26 | 4.48 | 3.45 | 4.44 | 3.85 | 4.47 | 3.98 | 4.52 | 4.26 | 4.18 |
| 1100 | 5.46 | 5.86 | 6.30 | 4.61 | 5.91 | 5.88 | 5.66 | 6.49 | 6.18 | 5.71 | 5.81 |
| 1200 | 6.04 | 6.71 | 7.48 | 6.39 | 6.83 | 8.11 | 7.13 | 8.52 | 8.22 | 7.46 | 7.29 |
| 1300 | 4.06 | 4.76 | 5.59 | 5.68 | 5.59 | 6.97 | 6.56 | 6.85 | 6.22 | 6.14 | 5.84 |
| 1400 | 1.19 | 1.78 | 2.41 | 3.24 | 3.07 | 3.28 | 3.47 | 2.88 | 2.36 | 2.84 | 2.65 |
| 1500 | 0.00 | 0.44 | 0.79 | 1.64 | 1.40 | 1.11 | 1.30 | 0.95 | 0.93 | 0.98 | 0.95 |
| 1600 | 0.00 | 0.33 | 0.09 | 1.04 | 0.83 | 0.53 | 0.88 | 0.61 | 0.69 | 0.43 | 0.54 |
| 1700 | 0.00 | 0.31 | -0.22 | 0.73 | 0.51 | 0.31 | 0.60 | 0.21 | 0.47 | 0.22 | 0.31 |
| 1800 | 0.00 | 0.26 | 0.10 | 0.43 | 0.21 | 0.22 | 0.25 | 0.10 | 0.30 | 0.17 | 0.20 |
| 1900 | 0.00 | 0.19 | 0.30 | 0.25 | 0.07 | 0.10 | 0.09 | 0.13 | 0.18 | 0.12 | 0.14 |

Table 3 shows the first derivative of the minimum ELC matrix with respect to the number of MERs. In general, the first derivative of minimum ELC continues to increase as the number of MERs increases. This implies that the rate of decrease in minimum ELCs is lower for higher number of MERs and it is not efficient to increase the number of MERs. The last row of Table 3 shows the vector of the average (over total sizes of MERs) of the first derivatives of the minimum ELCs. We have defined the optimal number of MERs as the number corresponding to the entry of the vector whose sign changes from negative to positive. Therefore, the optimal number of MERs for this case is seven.

Table 4 shows the second derivative of the minimum ELC matrix with respect to the total size of MERs. The second derivative of minimum ELC increases, reaches a maximum point, and then decreases as the total size of MERs is increased. The last column of Table 4 gives the average (over numbers of MERs) vector of the second derivatives of minimum ELCs, which is also shown in Figure 6. The average of the second derivative of minimum ECL is maximum when the total MER size is 1200 kW. Therefore, the optimal size of MERs for the case under investigation is 1200 kW.

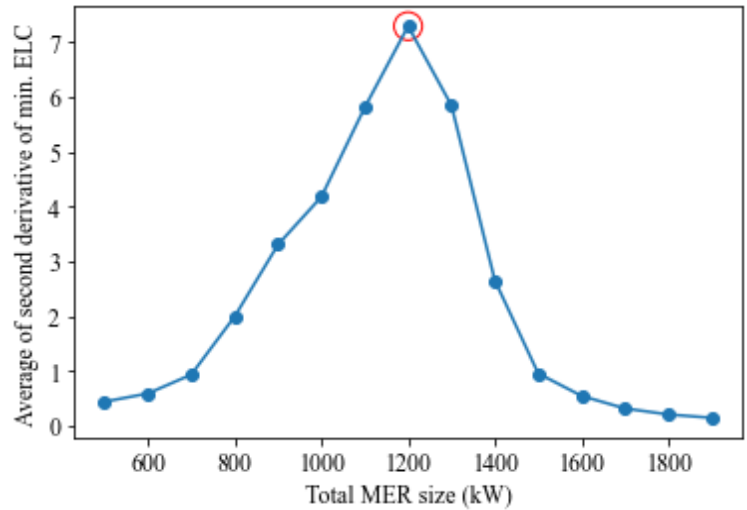

Figure 6: Plot of the average of second derivative of minimum ELC versus total MER size

## 5  CONCLUSION

This paper has proposed an approach based on graph theory and combinatorial enumeration to determine optimal size and number of movable energy resources (MERs) for enhanced electricity system resilience. Multiple line outage scenarios were generated, and the k-means method was used to reduce the generated scenarios to a manageable number of scenarios. The expected load curtailment (ELC) corresponding to each locational combination of MERs is determined considering the reduced line outage scenarios. The minimum ELC matrix was constructed to determine optimal size and number of MERs. The case study on a 33-node distribution system demonstrated the effectiveness of the proposed approach in determining the overall size and number of MERs for a spectrum of potential contingencies. Applying this research to real-world examples, using historic outage events, and to other types of distribution test systems would be valuable for future research.

# ACKNOWLEDGEMENT

This work was supported by the U.S. National Science Foundation (NSF) under Grant NSF 1847578.